\documentclass[12pt]{iopart}

\usepackage{iopams,graphicx,siunitx,braket,url}

\newcommand{\eb}{\epsilon_\text{b}}
\newcommand{\ed}{\epsilon_\text{b}^{(0)}}
\newcommand{\ea}{\epsilon_\text{b}^{(1)}}
\newcommand{\lmax}{l_\text{max}}

\newcommand{\rc}{r_\text{c}}

\begin{document}

\title{Effect of dipole polarizability on positron binding by strongly polar molecules}
\author{G F Gribakin and A R Swann}
\address{Centre for Theoretical Atomic, Molecular and Optical Physics,
School of Mathematics and Physics, Queen's University Belfast, Belfast BT7 1NN,
United Kingdom}
\ead{\mailto{g.gribakin@qub.ac.uk}, \mailto{aswann02@qub.ac.uk}}

\begin{abstract}
A model for positron binding to polar molecules is considered by combining the
dipole potential outside the molecule with a strongly repulsive core of a given
radius. Using existing experimental data on binding energies leads to
unphysically small core radii for all of the molecules studied. This suggests
that electron-positron correlations neglected in the simple model play a large
role in determining the binding energy. We account for these by including
polarization potential via perturbation theory and non-perturbatively. The
perturbative model makes reliable predictions of binding energies for a range
of polar organic molecules and hydrogen cyanide. The model also agrees with the
linear dependence of
the binding energies on the polarizability inferred from the experimental data
[Danielson \etal  2009 \textit{J.\ Phys.\ B: At.\ Mol.\ Opt.\ Phys.}\ {\bf 42}
235203]. The effective core radii, however, remain unphysically small for most
molecules. Treating molecular polarization non-perturbatively leads to
physically meaningful core radii for all of the molecules studied and enables
even more accurate predictions of binding energies to be made for nearly all
of the molecules considered.
\end{abstract}


\submitto{\jpb}

\section{Introduction}

Positrons are a useful tool in many areas of science, such as condensed matter 
physics, surface science and medicine (see, e.g., \cite{makkonen,wahl}).
Despite this, there is still much about their interactions with ordinary matter
that remains to be explored theoretically. In particular, the binding of
positrons to matter has been a difficult subject to research \cite{MBR02}. On
the part of theory, this is due to the strong electron-positron correlations
which
determine the binding energy and, in many cases, ensure the very existence of
bound states. On the experimental side, positron binding to atoms has not been
verified experimentally, largely due the difficulty in obtaining the relevant
species in the gas phase. On the other hand, for polyatomic molecules a wealth
of information is now available thanks to the special role that vibrational
Feshbach resonances play in positron-molecule annihilation 
\cite{gribakin}.

Before a positron annihilates with an electron in a molecule, it usually forms
a quasibound state with the molecule by transferring its excess energy into
vibrations of a single mode with near-resonant energy. This leads to pronounced
resonances observed in the positron-energy dependence of the annihilation
rate \cite{gribakin,gilbert}. Using the relation
\begin{equation}
\epsilon_\nu = \omega_\nu - \eb,
\end{equation}
where $\epsilon_\nu$ is the energy of the resonance due to vibrational 
mode $\nu$ with energy $\omega_\nu$, experimentalists have now been able to 
measure values of the positron binding energy $\eb$ for over sixty molecules 
\cite{danielson1,danielson3}. These measurements led to the construction of a 
phenomenological parametric fit of $\eb$ in terms of the the dipole
polarizability $\alpha$ and permanent dipole moment $\mu$ of the molecule:
\begin{equation}\label{eqn:regression}
\eb = 12.4 (\alpha + 1.6 \mu - 5.6),
\end{equation}
where $\eb$ is in milli-electron volts, $\alpha$ is in cubic angstroms and $\mu$
is in debyes (D) \cite{danielson4}. An interesting feature of
\eref{eqn:regression} is that the dependences of $\eb$ on $\mu$ and $\alpha$ are
both linear. Although a general increase of $\eb$ with $\mu$ and $\alpha$ is to
be expected (since both contribute to the positron-molecule attraction), there
is no obvious reason why the dependences should be linear. In fact,
measurements for some molecules with large dipole moments, such as acetone and
acetonitrile, yielded binding energies more than double the values predicted
by equation \eref{eqn:regression} \cite{danielson1}.

Despite the wealth of experimental data on positron-molecule binding energies,
theoretical developments are somewhat behind. There are few calculations of
positron binding to nonpolar or weakly polar molecules. The zero-range
potential model \cite{Gribakin06,Gribakin09} captured the qualitative features
of the binding for alkanes and correctly predicted the emergence of the second
bound state \cite{Gribakin06}. There were also predictions of positron binding
to the hydrogen molecule in the excited $A$~$^3\Sigma _u$ state
\cite{PhysRevA.83.064701}, and configuration interaction calculations for
carbon-containing triatomics (CO$_2$, CS$_2$, CSe$_2$ and weakly polar COS,
COSe and CSSe) \cite{Koyanagi2013a,Koyanagi2013b}. The latter papers
reported binding by the two heaviest species in the vibrational ground state,
by CS$_2$ in the lowest vibrationally excited states, and by other molecules at
higher vibrational excitations or upon bond deformations. In contrast, there is
a large number of quantum chemistry calculations of positron binding with
strongly polar polyatomic molecules with dipole moments $\gtrsim \SI{3}{D}$.
For such molecules binding is achieved even at the lowest, static-field (e.g.,
Hartree-Fock) level of theory. The static-field binding energies, however, are
usually quite small, and the effect of correlations (e.g., polarization of the
molecule by the positron) increases the binding energy dramatically (see, e.g.,
\cite{tachikawa1,strasburger,chojnacki,kita,Romero2014}). Recent configuration
interaction calculations for nitriles, acetaldehyde, and acetone 
\cite{tachikawa2,Tachikawa2012,Tachikawa2014} in fact give binding energies
within 25--50\% of the experiment, which is quite good, given the complexity
of the system.

\begin{figure}[ht!]
\begin{indented}
\item[]\includegraphics[width=.8\textwidth,clip=true]{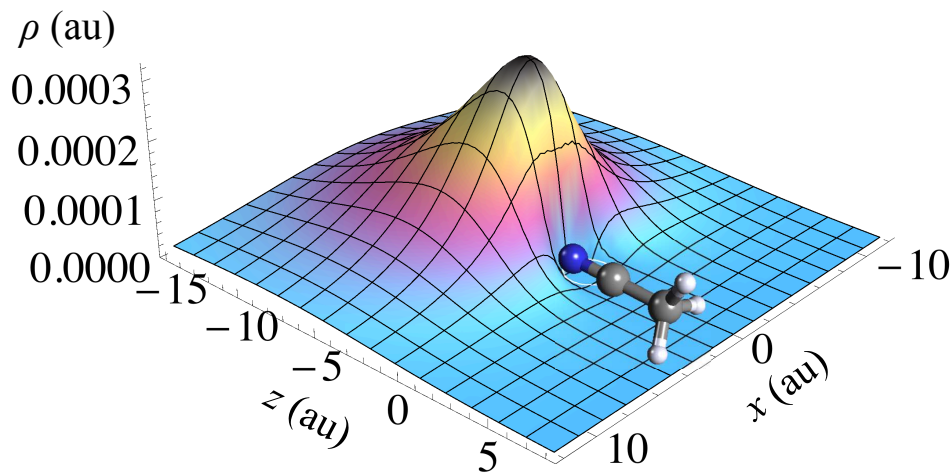}
\caption{The density of the positron bound in the field of a point dipole with
the dipole moment $\mu=\SI{3.93}{D}$ of the acetonitrile molecule and repulsive core of
the radius $r_0=\SI{1.175}{au}$, with the binding energy of $\eb =\SI{27}{meV}$.}
\label{fig:ch3cn}
\end{indented}
\end{figure}

The purpose of this article is to present a simple model for positron binding 
to polar molecules. For many molecules of interest the dipole moment is
dominated by a single bond (e.g., CN or CO), located at one end of the molecule,
with the negative charge on the terminal atom. Given the positron repulsion
from the atomic nuclei, we model the molecular potential as a point dipole
surrounded by an impenetrable sphere. Of course, the true size of the
molecular dipole is finite, of the order of interatomic distances. However, for
weakly bound positron states, the wave function of the positron is very diffuse.
Its spatial extent is much greater than the physical size of the dipole, which
justifies the applicability of the point-dipole model to weakly bound
positron states. This is illustrated by figure \ref{fig:ch3cn}, which shows the
density for the positron bound in the dipole field of the acetonitrile molecule
(CH$_3$CN). The figure also shows that the positron is localized in the
negative-energy well of the dipole potential and is largely ``unaware'' of the
true geometry of the molecule. This justifies the hard-sphere model for the
short-range positron repulsion. Quantum chemistry calculations of the positron
density in polar molecules support the picture of a diffuse positronic cloud
localized off the negatively charged end of the molecular dipole
\cite{tachikawa1,kita,tachikawa2,Tachikawa2012}. 

Note that a recent paper \cite{PhysRevA.86.042708} combined a hard-sphere
repulsive core with the polarization potential to model positron binding to
atoms and nonpolar molecules. While the two models bear some similarity, the
physics of positron binding to neutral atoms and nonpolar species is very
different from that of binding to strongly polar molecules explored in this
work. In the former case, for atoms the positron does form a spherical cloud,
but for molecules the shape of the positron wave function largely repeats that
of the molecule \cite{Gribakin09}, and the spherical repulsive core
approximation is hard to justify. In contrast, for bound states with polar
species, the positron resides in the dipole-generated well to one side of the
molecule (see figure \ref{fig:ch3cn}), and the repulsive core model looks more
appropriate.

The main features of binding by the dipole potential are outlined in section
\ref{sec:theory}. Applying the model to polar molecules for which the positron
binding energies are known from experiment (section \ref{sec:dipole}) shows
that electron-positron correlations have a large effect on binding. We include
these in the form of the polarization potential, first via perturbation theory
(section \ref{sec:perturb}) and then non-perturbatively
(section \ref{sec:nonpert}). While this model may appear to be rather crude, it
captures the main physical aspects of the problem. Owing to its simplicity, it
provides a deeper understanding of some of the key features that have been
observed in experiment, including the empirical scaling (\ref{eqn:regression}).
The usefulness of such models as a means of obtaining an \textit{explanation},
and complementary to heavy numerical computations, was argued well by
Ostrovsky, who introduced the notion of complementarity between calculation
and explanation \cite{ostrovsky1,ostrovsky2}.

\section{Theory}\label{sec:theory}

We model the molecule as an impenetrable sphere of radius $r_0$, with a point 
dipole of dipole moment $\bmu$ fixed at its centre (the origin). The
positron experiences point-dipole potential in the region outside the sphere.
Using spherical polar coordinates $(r,\theta,\phi)$ and choosing the polar
($z$-) axis along $ \bmu$, we have
\begin{equation}\label{eqn:potential}
V_\mathrm{d}(\mathbf{r}) = \cases{\infty &\mbox{for $r\leq r_0$},\\ \mu r^{-2}\cos\theta  
& \mbox{for $r>r_0$},}
\end{equation}
where $\theta$ is the polar angle, and we work in atomic units (au).

Although \eref{eqn:potential} is a non-central potential, the Schr\"odinger
equation,
\begin{equation}\label{eqn:schrodinger}
\left[ -\frac{1}{2} \nabla^2 + V_\mathrm{d}(\mathbf{r}) \right]
\psi(\mathbf{r}) = E \psi({\bf r}),
\end{equation}
for the positron wave function $\psi({\bf r})$ and energy $E$ can be 
solved in the region $r>r_0$ using separation of variables. Inserting the
ansatz $\psi({\bf r}) = R(r)\Phi(\theta,\phi)$ into \eref{eqn:schrodinger}
yields separate radial and angular equations:
\numparts
\begin{eqnarray}
\frac{1}{r^2} \frac{\rmd}{\rmd r} \left( r^2 \frac{\rmd R}{\rmd r} \right) + \left( 2E - 
\frac{\lambda}{r^2} \right)R = 0,\label{eqn:radial}\\
\frac{1}{\sin\theta} \frac{\partial}{\partial\theta} \left( \sin\theta \, \frac{\partial\Phi}
{\partial\theta} \right) + \frac{1}{\sin^2\theta} \frac{\partial^2 \Phi}{\partial\phi^2} + 
(\lambda - 2\mu\cos\theta)\Phi = 0,\label{eqn:angular}
\end{eqnarray}
\endnumparts
where $\lambda$ is a separation constant. If $\mu=0$ then \eref{eqn:angular} 
becomes $(\hat{L}^2 - \lambda)\Phi=0$, where $\hat{L}^2$ is the squared angular 
momentum operator. This is just the eigenvalue equation for the $\hat{L}^2$ 
operator; the possible values of $\lambda$ are $l(l+1)$, where $l$ (the azimuthal 
quantum number) is a non-negative integer, and the eigenfunctions are the 
spherical harmonics $Y_{lm}(\theta,\phi)$, where $m$ (the magnetic quantum 
number) is the eigenvalue of $\hat{L}_z$, and $m$ is an integer, $|m| \leq l$.

For $\mu \neq 0$, $l$ is no longer a good quantum number since $\hat{L}^2$ does 
not commute with the Hamiltonian. However, $\hat{L}_z$ does commute with the 
Hamiltonian, and thus $m$ is still a good quantum number. We must solve the 
angular equation \eref{eqn:angular} to find the new values of $\lambda$. With this 
information we will be able to solve the radial equation and use it to investigate the 
dependence of the binding energy $\eb =|E|$ on $\mu$ and $r_0$.

\subsection{Angular equation}

Since $m$ is a good quantum number, there will be a distinct set of
eigenfunctions $\Phi_m(\theta,\phi)$ of the angular equation \eref{eqn:angular}
for each value of $m$. We expand the unknown functions in the basis of
spherical harmonics, i.e.,
\begin{equation}\label{eqn:angularexpansion}
\Phi_m(\theta,\phi) = \sum_{l'=|m|}^\infty C_{l'm} Y_{l'm}(\theta,\phi)
\qquad (m=0,\,
\pm1,\,\pm2,\dots),
\end{equation}
where the $C_{l'm}$ are unknown numbers. Substituting this expression into 
\eref{eqn:angular}, multiplying across by $Y_{lm}^*(\theta,\phi)$, where $l$ is
a non-negative integer such that $l\geq |m|$, integrating over $\phi $ and
$\theta $ and using properties of spherical harmonics (see, e.g.,
\cite{varshalovich}) yields
\begin{equation}\label{eqn:matrixeigenvalue}
\sum_{l'=|m|}^\infty B_{ll'm} C_{l'm} = \lambda_m C_{lm},
\end{equation}
where
\begin{eqnarray}
\fl B_{ll'm} = l(l+1)\delta_{ll'}
+ 2\mu(-1)^m \sqrt{(2l+1)(2l'+1)} \left({1\atop 0}~{l\atop 0}~{l'\atop 0}\right)
\left( {1\atop 0}~{l\atop -m}~{l'\atop m}\right),
\end{eqnarray}
and the arrays in parentheses are $3j$ symbols. The eigenvalue $\lambda$ has 
been renamed $\lambda_m$ since it will have different sets of values depending
on $m$. Equations \eref{eqn:matrixeigenvalue} are a set of matrix eigenvalue 
equations for the semi-infinite, symmetric, tridiagonal matrix $B_{ll'm}$,
whose rows and columns are enumerated by $l$ and $l'$, respectively. Each of
these matrices has a countably infinite set of eigenvalues, so we rename
$\lambda_m$ as $\lambda_{ms}$, where $s=1,\,2,\,3,\dots$ enumerates the
different eigenvalues for each $m$, and the eigenvalues are arranged so that
$\lambda_{m1} <\lambda_{m2} <\lambda_{m3}<\dots $. Symmetry properties of the
$3j$ symbols can easily be used to show that $B_{ll',-m} = B_{ll'm}$, and so 
\eref{eqn:matrixeigenvalue} need only be solved for $m\geq 0$.

\begin{figure}[ht!]
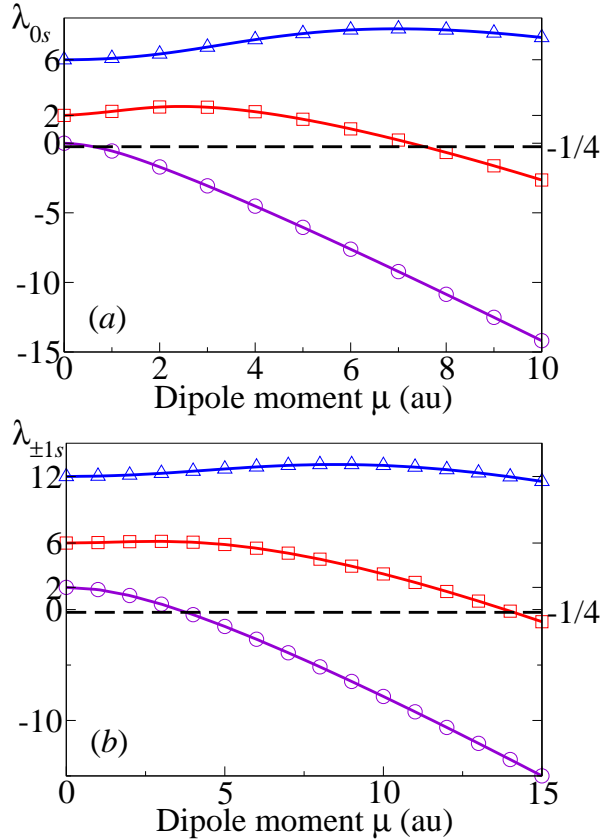

\begin{indented}
\item[]\includegraphics[width=.5\textwidth,clip=true]{figure2a.eps}\\
\includegraphics[width=.5\textwidth,clip=true]{figure2b.eps}
\caption{\label{fig:eigenvalues}Dependence of the eigenvalues $\lambda_{m1}$ 
(purple circles), $\lambda_{m2}$ (red squares) and $\lambda_{m3}$ (blue
triangles) for (\emph{a}) $m=0$ and (\emph{b}) $m=\pm1$, on the dipole moment
$\mu$. Intersections with the dashed lines ($\lambda =-1/4$) give critical dipole
moments $\mu_{\rm crit}$.}
\end{indented}
\end{figure}

We seek bound states of the positron. From the form of the radial equation 
\eref{eqn:radial} it can be shown that for $\lambda_{ms} < -\frac{1}{4}$ there
will be an infinite number of bound states, while for
$\lambda_{ms} > -\frac{1}{4}$ there will be none \cite{landau} (see section
\ref{subsec:radeq}). Given a certain value of $\mu$ and of $m$, by truncating the
infinite matrix $B_{ll'm}$ to a finite size (where the final row and column are
denoted by $l=l'=\lmax$), we can find numerical approximations for the first
$\lmax - |m|+1$ values of $\lambda_{ms}$. Table \ref{tab:eigenvalues} and
figure \ref{fig:eigenvalues} show how $\lambda_{m1}$, $\lambda_{m2}$ and
$\lambda_{m3}$ vary with $\mu$ for $m=0$ and $m=\pm 1$. Values of $\lmax $ shown
in the last column of table  \ref{tab:eigenvalues} are chosen so that the
eigenvalues are correct to at least six decimal places.

\begin{table}[ht!]
\caption{\label{tab:eigenvalues} Values of $\lambda_{ms}$ for $m=0,\,\pm1$, 
$s=1\,,2,\,3$ across a range of values of the dipole moment $\mu$. The values of $
\lmax$ needed for stability to six decimal places are also shown.}
\lineup
\begin{indented}
\item[]\begin{tabular}{@{}llllll}
\br
$|m|$ & $\mu$ (au) & $\lambda_{m1}$ & $\lambda_{m2}$ & $\lambda_{m3}$ & $
\lmax$ \\
\mr
0 & \00 & \00.000\,000 & 2.000\,000 & \06.000\,000 & \02 \\
0 & \02 & \-\01.704\,857 & 2.602\,337 & \06.412\,828 & \07 \\
0 & \04 & \-\04.519\,910 & 2.263\,955 & \07.444\,429 & \08 \\
0 & \06 & \-\07.616\,374 & 1.031\,162 & \08.141\,444 & \09 \\
0 & \08 & \-10.856\,049 & \-0.662\,826 & \08.138\,189 & 10 \\
0 & 10 & \-14.186\,766 & \-2.640\,671 & \07.597\,027 & 10 \\
1 & \00 & \02.000\,000 & 6.000\,000 & 12.000\,000 & \03 \\
1 & \03 & \00.489\,539 & 6.153\,928 & 12.285\,114 & \08 \\
1 & \06 & \-\02.675\,243 & 5.535\,499 & 12.865\,402 & \09 \\
1 & \09 & \-\06.481\,474 & 3.913\,801 & 13.078\,672 & 10 \\
1 & 12 & \-10.628\,924 & 1.630\,021 & 12.620\,576 & 11 \\
1 & 15 & \-14.995\,686 & \-1.093\,056 & 11.558\,712 & 12 \\
\br
\end{tabular}
\end{indented}
\end{table}

Considering the eigenvalues $\lambda_{ms}$ as functions of $\mu$, for each 
combination of $m$ and $s$ there is a critical dipole moment $\mu_{\rm crit}$
for which $\lambda_{ms}=-\frac{1}{4}$. Some of these critical dipole
moments are shown in table \ref{tab:critical}; they agree with the values
obtained by Fermi and Teller \cite{fermi} and Crawford \cite{crawford}. The
condition $\mu>\mu_{\rm crit}$ guarantees binding by either a point-like or finite dipole.
The smallest critical dipole is $\mu_{\rm crit} = \SI{1.625}{D}$ for $m=0$, $s=1$.
\begin{table}[ht!]
\caption{\label{tab:critical}Critical dipole moments required for various bound
states of the positron, along with the values of $\lmax$ required for stability
to \SI{e-6}
{au}.}
\lineup
\begin{indented}
\item[]\begin{tabular}{@{}lllll}
\br
$|m|$ & $s$ & $\mu_{\rm crit} $ (au) & $\mu_{\rm crit} $ (D) & $\lmax$ \\
\mr
0 & 1 &  \00.639\,315 &  \01.625 & \04 \\
0 & 2 &  \07.546\,956 &  19.182 & 10 \\
1 & 1 &  \03.791\,968 &  \09.634 & \08 \\
1 & 2 & 14.112\,115 & 35.869 & 12 \\
2 & 1 &  \09.529\,027 & 24.220 & 10 \\
\br
\end{tabular}
\end{indented}
\end{table}
Since typical molecules have dipole moments not exceeding \SI{11}{D}, i.e.,
up to \SI{4.3}{au}, the only possible bound states are those corresponding to
$m=0$, $s=1$ and $m=\pm1$, $s=1$. The critical value of $\mu$
needed to sustain any other bound state is simply too high.

It must be mentioned that the above considerations apply to binding by the
static dipole, i.e., assuming that the molecules cannot rotate. When rotations
are included, the values of $\mu_{\rm crit} $ required for the dipole binding to occur
are 10-30\% greater \cite{Garrett71}. This gap depends on the moment of inertia
of the molecule and increases with the molecular angular momentum, being
smallest for large, slowly rotating molecules. Another consideration important
for \textit{ab initio} quantum-chemistry calculatons of binding is that for the
values of $\mu$ only slightly exceeding $\mu_{\rm crit}$, the binding energy is very
sensitive to the actual value of the dipole moment (see figure
\ref{fig:energyvsD} in section \ref{subsec:BEr_0D}). The actual value of
the dipole moment depends on the approximation used (e.g., Hartree-Fock), and
can be a significant source of error \cite{CG77}. However, both the effect of
molecular rotations and the sensitivity to the value of $\mu$ are offset by the
large contribution of electron-positron correlations to the positron binding
(see sections \ref{sec:perturb} and \ref{sec:nonpert}).

\subsection{Radial equation}\label{subsec:radeq}

Under the substitution
\begin{equation}\label{eqn:substitution}
R(r) = \frac{Z(kr)}{\sqrt{kr}},
\end{equation}
where $k=\sqrt{2E}$, the radial equation \eref{eqn:radial} yields the following 
differential equation for $Z(kr)$:
\begin{equation}
(kr)^2 \frac{\rmd^2 Z}{\rmd (kr)^2} + kr \frac{\rmd Z}{\rmd (kr)} + \left[ (kr)^2 - 
\left( \lambda_{ms} + \frac{1}{4} \right) \right] = 0.
\end{equation}
This is just Bessel's differential equation. Since for bound states we have $E < 0$, 
i.e., $E=-\vert E\vert$, it is best to express the general solution in terms of modified 
Bessel functions:
\begin{equation}
Z_{ms}(kr) = A_{ms} K_{\rmi \beta_{ms}} (\kappa r) + B_{ms} I_{\rmi \beta_{ms}} 
( \kappa r),
\end{equation}
where the subscripts $m$ and $s$ have been added to $Z$ because there is a 
distinct function for each combination of $m$ and $s$, $A_{ms}$ and $B_{ms}$ are 
arbitrary constants, $\kappa=\sqrt{2|E|}$, and
\begin{equation}\label{eq:beta}
\beta_{ms} = \left|\lambda_{ms} + \frac{1}{4}\right|^{1/2},
\end{equation}
and $\lambda_{ms} <-\frac{1}{4}$ is assumed. Equation \eref{eqn:substitution} then 
gives
\begin{equation}
R_{ms}(r) = A_{ms} \frac{K_{\rmi \beta_{ms}} (\kappa r)}{\sqrt{\kappa r}} + B_{ms} 
\frac{I_{\rmi \beta_{ms}} ( \kappa r)}{\sqrt{\kappa r}}.
\end{equation}

For the bound-state wave function to be normalizable we must require $R_{ms}(r) 
\rightarrow 0$ as $r \rightarrow \infty$. It can be seen from the asymptotic forms of 
the modified Bessel functions (see, e.g., \cite{stegun}) that $I_\nu (x) / \sqrt{x} 
\rightarrow \infty$ as $x \rightarrow \infty$, while $K_\nu (x) / \sqrt{x} \rightarrow 0$ 
as $x \rightarrow \infty$, assuming that $x$ is real. We therefore require $B_{ms} = 
0$ for every $m$ and $s$, and so
\begin{equation}\label{eq:Rms}
R_{ms}(r) = A_{ms} \frac{K_{\rmi \beta_{ms}} (\kappa r)}{\sqrt{\kappa r}},
\end{equation}
with $A_{ms}$ a constant of normalization.

Since $\kappa r$ and $\beta_{ms}$ are real and positive, the function $K_{\rmi
\beta_{ms}} (\kappa r)$ (also known as the Macdonald function) is also real, which 
can be seen, e.g., from the integral representation \cite{watson},
\begin{equation}
K_{\rmi \beta_{ms}} (\kappa r) = \int_0^\infty \exp (-\kappa r \cosh t) \cos (\beta_{ms} 
t) \, \rmd t,
\end{equation}
and thus $R_{ms}(r)$ is real (for a real $A_{ms}$). The function $R_{ms}(r)$ has 
infinitely many positive roots, with an accumulation point at $r=0$.

The second boundary condition to be applied to $R_{ms}(r)$ is due to the 
impenetrable sphere at $r=r_0$, which means that we must have $R_{ms}(r_0)=0$, 
i.e.,
\begin{equation}
\frac{K_{\rmi\beta_{ms}}(\kappa r_0)}{\sqrt{\kappa r_0}}= 0.
\end{equation}
The positive roots of the function $K_{\rmi\beta_{ms}}(z)$ are therefore the allowed 
values of $\kappa r_0$. Since these roots form an infinite, discrete set, we shall 
name them $\zeta_{msn}$, where $n=1,\,2,\,3,\dots$, and $\zeta_{ms1} > 
\zeta_{ms2} > \zeta_{ms3} > \dots$. These roots can be found numerically. For any 
particular molecule, $r_0$ is a constant, and so the permissible values of $\kappa$ 
(which we now rename $\kappa_{msn}$, and likewise with $E$) are $\kappa_{msn} 
= \zeta_{msn}/r_0$, i.e., 
\begin{equation}
E_{msn} = -\frac{\kappa_{msn}^2}{2} = - \frac{1}{2} \left( \frac{\zeta_{msn}}{r_0} 
\right)^2.
\end{equation}

\subsection{Dependence of binding energy on $r_0$ and $\mu$}\label{subsec:BEr_0D}

\begin{figure}
\begin{indented}
\item[]\includegraphics*[width=.6\textwidth]{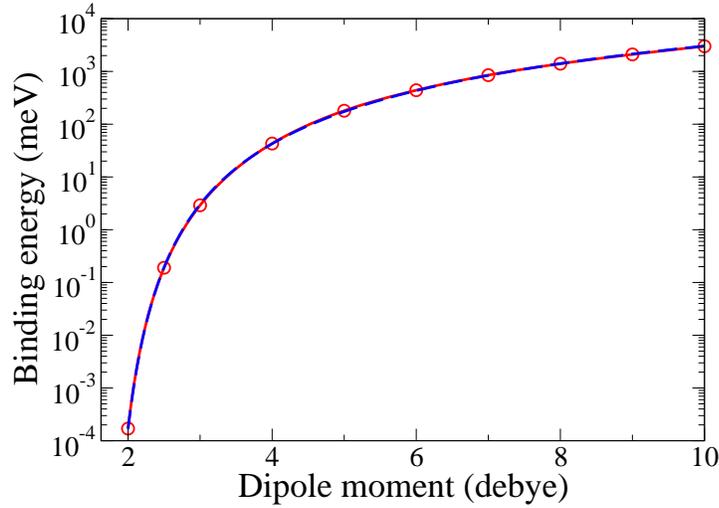}
\caption{Dependence of the ground-state binding energy
$\eb$ on $\mu$ for $r_0=1$. Solid red line with circles shows the numerical
results, and the blue dashed line is the fit (\ref{eq:BE_D}) with
$A=\SI{65998.7}{meV}$, $B=\SI{12.2705}{D^{1/2}}$, and $C=\SI{0.400612}
{D^{-1/2}}$. Note that the two curves are indistinguishable on the scale
of the graph.}
\label{fig:energyvsD}
\end{indented}
\end{figure}

For a given value of the dipole moment, the largest negative value of
$\lambda _{ms}$ is for $m=0$, $s=1$, with the critical
dipole moment $\mu_{\rm crit} =\SI{0.6393}{au}=\SI{1.625}{D}$
\cite{fermi,crawford}. The corresponding ground-state binding energy is
\begin{equation}\label{eq:BE}
\eb = \frac{1}{2} \left( \frac{\zeta_{011}}{r_0} \right)^2,
\end{equation}
where $\zeta_{011}$ is the largest root of $K_{\rmi\beta_{01}}(z)$, whose index
$\beta _{01}$ is determined by the eigenvalue $\lambda _{01}$ of the angular
equation, see \eref{eq:beta}.
The dependence of the binding energy \eref{eq:BE} on $r_0$ is simple. Figure 
\ref{fig:energyvsD} shows the dependence of $\eb $ on the magnitude of the
dipole moment $\mu$ for the fixed repulsive core radius $r_0=\SI{1}{au}$
(i.e., $\eb = \frac{1} {2} \zeta_{011}^2$).

For $\mu\rightarrow \mu_{\rm crit} $, the binding energy rapidly tends to zero. This limit
corresponds to $\lambda _{01}\rightarrow -\frac{1}{4}$ and
$\beta _{01}\rightarrow 0$. In the limit of small $\beta $, the
roots of the Macdonald function $K_{\rmi\beta }(z)$ have the following asymptotic
behaviour \cite{ferreira}:
\begin{equation}\label{eq:root}
\ln z_n\simeq -\frac{n\pi }{\beta }+\ln 2-\gamma \qquad (n=1,\,2,\dots ),
\end{equation}
where $\gamma \approx 0.577$ is Euler's constant. The largest root that we are
interested in ($n=1$) is then given by
\begin{equation}\label{eq:zeta011}
\zeta _{011}\simeq 2\rme^{-\gamma}\exp \left(-\frac{\pi }{\beta _{01}}\right). 
\end{equation}
For dipole moments close to the critical value, we have from equation 
\eref{eq:beta},
\begin{equation}\label{eq:beta_as}
\beta _{01}\simeq \left[\left.-\frac{\rmd\lambda _{01}}{\rmd \mu}
\right|_{\mu=\mu_{\rm crit} }(\mu-\mu_{\rm crit} )\right]^{1/2}.
\end{equation}
Combining equations \eref{eq:BE}, \eref{eq:zeta011} and \eref{eq:beta_as}, gives
\begin{equation}\label{eq:lead}
\eb =A \exp \left[-B(\mu-\mu_{\rm crit})^{-1/2}\right],
\end{equation}
where $A$ and $B$ are constants (see also \cite{Abramov72,Fabrikant83}, from
which a similar result can be derived). Motivated by this scaling,
we constructed an approximate analytical expression for the binding energy as
a function of $\mu$ in the following form:
\begin{equation}\label{eq:BE_D}
\eb =A \exp \left[-B(\mu-\mu_{\rm crit} )^{-1/2}+C(\mu-\mu_{\rm crit} )^{1/2}\right].
\end{equation}
Here the second term in the exponent represents a correction to the leading
term (\ref{eq:lead}). It accounts for the next order corrections in both
\eref{eq:zeta011} and \eref{eq:beta_as}, and extends the applicability of
\eref{eq:BE_D} way beyond
the range of near-critical $\mu$. Regarding the constants $A$, $B$ and $C$ as
fitting parameters, an excellent fit of the numerical data over the
whole range covered by figure \ref{fig:energyvsD} is obtained using $A=
\SI{65998.7}{meV}$, $B=\SI{12.2705}{D^{1/2}}$ and $C=\SI{0.400612}{D^{-1/2}}$, 
and the dipole moment $\mu$ in debye (D).

\section{Positron binding by the dipole potential}\label{sec:dipole}

Using experimental data on positron binding energies from
\cite{danielson3,young} and dipole moments from \cite{CRC}, we fitted the
energies to equation \eref{eq:BE} by adjusting the values of $r_0$ for fourteen
polar organic molecules and three polar inorganic molecules\footnote{The methyl
halide molecules are quite distinct from the other molecules studied. Each of
them contains a different halogen atom. It is this, rather than the size of the
molecule, that affects their dipole polarizability.}. No direct experimental
binding energy is available for methyl fluoride ($\mathrm{CH_3F}$), and the
value obtained by fitting theoretical annihilation rate to experiment has been
used instead \cite{GribakinLee06}. If the static dipole potential provided the 
dominant contribution to the binding, then it could be expected that for
molecules with a single dipolar bond the values of $r_0$ would be approximately
half the length of the molecular dipole (${\sim}\SI{1}{au}$), with values
significantly larger or smaller than this considered as unphysical. For
molecules with several dipolar bonds (e.g., the formates and acetates) we
expected a larger value of $r_0$ than in the case of a single dipole bond.

The results are shown in table \ref{tab:moleculedata}.
\begin{table}
\caption{\label{tab:moleculedata} Values of $r_0$ obtained for a selection of 
molecules by fitting known binding energies from 
\cite{danielson3,young,GribakinLee06} to equation \eref{eq:BE}.}
\lineup
\begin{indented}
\item[]\begin{tabular}{@{}llll}
\br
Molecule & $\mu$ (D) & $\eb$ (meV) &$r_0$ (au) \\
\mr
\bf{Aldehydes} \\
Acetaldehyde ($\mathrm{C_2 H_4 O}$) & 2.75  & \088 & \num{1.03e-1} \\
Propanal ($\mathrm{C_3 H_6 O}$)& 2.52  & 118 & \num{4.28e-2} \\
Butanal ($\mathrm{C_4 H_8 O}$)& 2.72  & 142 & \num{7.44e-2} \\
\bf{Ketones} \\
Acetone ($\mathrm{C_3 H_6 O}$)& 2.88  & 174 & \num{1.00e-1} \\
2-butanone ($\mathrm{C_4 H_8 O}$) & 2.78  & 194 & \num{7.45e-2} \\
Cyclopentanone ($\mathrm{C_5 H_8 O}$) & 3.30  & 230 & \num{1.90e-1} \\
\bf{Formates} \\
Methyl formate ($\mathrm{C_2 H_4 O_2}$) & 1.77  & \065 & \num{4.04e-6} \\
Ethyl formate ($\mathrm{C_3 H_6 O_2}$)& 1.98  & 103 & \num{9.76e-3} \\
Propyl formate ($\mathrm{C_4 H_8 O_2}$)& 1.89  & 126 & \num{1.77e-4} \\
\bf{Acetates} \\
Methyl acetate ($\mathrm{C_3 H_6 O_2}$)& 1.72  & 122 & \num{7.38e-8} \\
Ethyl acetate ($\mathrm{C_4 H_8 O_2}$)& 1.78  & 160 & \num{4.32e-6}\\
\bf{Nitriles} \\
Acetonitrile ($\mathrm{C_2 H_3 N}$)& 3.93 & 180 & \num{4.54e-1} \\
Propionitrile ($\mathrm{C_3 H_5 N}$) & 4.05  & 245 & \num{4.37e-1} \\
2-methylpropionitrile ($\mathrm{C_4 H_7 N}$) & 4.29  & 274 & \num{5.04e-1} \\
\bf{Methyl halides} \\
Methyl fluoride ($\mathrm{CH_3 F}$) & 1.86 & 0.3 & \num{1.72e-3} \\
Methyl chloride ($\mathrm{CH_3 Cl}$) & 1.90 & 25 & \num{1.68e-5} \\
Methyl bromide ($\mathrm{CH_3 Br}$) & 1.82 & 40 & \num{1.67e-6} \\
\br
\end{tabular}
\end{indented}
\end{table}
Clearly, all of the radii obtained are unphysically small, particularly for the
most weakly polar molecules. The largest value, $r_0=\SI{0.5}{au}$, is for
the most strongly polar molecule studied: 2-methylpropionitrile, but even this
is less than a quarter of the C$\equiv$N bond length. 

It can be seen from table \ref{tab:moleculedata} that despite molecules of the
same type (aldehyde, ketone, etc.)\ having similar dipole moments, there can be 
significant variations in the binding energies. For example, consider the
molecules acetaldehyde and butanal. Their dipole moments are very close
(\SI{2.75}{D} and \SI{2.72}{D}, respectively), and the dipole in both molecules
is due to a C=O bond. Yet there is a large difference in the
binding energies: the binding energy for butanal (\SI{142}{meV}) is more than
1.5 times that for acetaldehyde (\SI{88}{meV}). Peculiarly, acetaldehyde is
actually slightly more polar than butanal, yet has the lower binding energy. A
similar situation also occurs with acetone and 2-butanone, and with ethyl
formate and propyl formate. These observations cannot be explained by our model
as it stands.

The fact that the values of $r_0$ obtained for all of the molecules studied are 
unphysically small implies that lepton correlations (in particular, due to
polarization of the molecule by the positron) play an important role in
enhancing the binding energy, even for strongly polar molecules. Going back to
the example of acetaldehyde vs butanal, acetaldehyde has a polarizability of
\SI{4.6}{\angstrom^3}, while the polarizability of butanal is a significantly
greater value of \SI{8.2} {\angstrom^3} \cite{CRC}. This explains the larger
binding energy of the latter molecule. In the following two sections we
investigate the effect of the molecular polarization on positron binding, and
show that its inclusion is critical for obtaining a correct physical picture of
positron binding to polar molecules.

\section{Perturbative correction due to molecular polarization}\label{sec:perturb}

\subsection{Core radii for perturbative inclusion of polarization}

Since the dipole potential for $\mu>\mu_{\rm crit} $ is sufficient to create a
``zeroth-order''  bound state, we first estimate the effect of molecular
polarization using perturbation theory. The values of the radius $r_0$ can then
be chosen by fitting the total binding energy (i.e., due to the dipole force
and polarization) to the experimental values, expecting that this should lead
to more realistic values of $r_0$.

The extra contribution to the positron potential energy (in the region $r>r_0$)
due to molecular polarization can be approximated by the polarization potential
\begin{equation}\label{eq:Vpol}
V_{\text{pol}} ({\bf r}) = -\frac{\alpha}{2r^4},
\end{equation}
where $\alpha $ is the molecular dipole polarizability. Using perturbation
theory, the first-order correction to the original dipole binding energy
(\ref{eq:BE}), which we now label $\eb^{(0)}$, is
\begin{equation}
\eb^{(1)} = \int \frac{\alpha}{2r^4} |\psi_{011}({\bf r})|^2 \, \rmd^3 {\bf r},
\end{equation}
where $\psi_{msn}({\bf r})=R_{msn}(r)\Phi _{ms}(\theta ,\phi )$. Assuming that
the radial and angular parts of the wave function are separately normalized to
unity, this becomes
\begin{equation}\label{eqn:correction}
\eb^{(1)} = \frac{\alpha}{2} \int_{r_0}^\infty |R_{011}(r)|^2r^{-2} \, \rmd r.
\end{equation}

For each of the molecules studied we made an initial estimate for the value of 
$r_0$ and adjusted it until the new binding energy $\eb = \eb^{(0)} + \eb^{(1)}$
was within 10\% of the experimental value (with both $\eb^{(0)}$ and $\eb^{(1)}$
being functions of $r_0$). The results are shown in table
\ref{tab:perturbeddata}. Polarizabilities are taken from the CRC Handbook
\cite{CRC}, with the exceptions of propyl formate and cyclopentanone, for which
the polarizabilities have been estimated by Danielson \etal \cite{danielson3}.

\begin{table}
\caption{\label{tab:perturbeddata} Fitted values of $r_0$ with the inclusion of 
polarization via perturbation theory. Also shown are the corresponding values
of $ \eb^{(0)}$ and $\eb^{(1)}$, the expectation values of the potential energy
due to the permanent dipole, and the predicted and experimental values of
$\eb$.}
\lineup
\begin{tabular}{@{}lllllllll}
\br
&&&&&&& \centre{2}{$\eb$ (meV)}  \\ \ns 
& $\mu$ &$\alpha$  & $r_0$  & $\eb^{(0)}$  &$\eb^{(1)}$ &
$\vert \langle V_\mathrm{d}({\bf r}) \rangle \vert$ & \crule{2}\\
Molecule & (D) & (\si{\angstrom^3}) & (au) & (meV) & (meV)&(meV)&Pred.& Exp.\\
\mr
\bf{Aldehydes} \\
Acetaldehyde & 2.75 & 4.6&0.60 &\03&\080&\038&\083&\088   \\
Propanal & 2.52  &6.5 &0.42&\01&118&\023&119&118  \\
Butanal & 2.72  &8.2&0.58&\02&140& \035&142&142   \\
\bf{Ketones} \\
Acetone & 2.88  &6.4 &0.63&\04&168&\058&172&174   \\
2-butanone  & 2.78  & 8.1&0.58&\03&187&\046&190&194   \\
Cyclopentanone  & 3.30  & 9.0&0.92&10&220&\098&230&230  \\
\bf{Formates} \\
Methyl formate & 1.77  &5.1 &0.004& ${\sim}\num{e-5}$ &\069&${\sim}\num{e-2}$&
\069&\065   \\
Ethyl formate & 1.98  &6.9 &0.066&${\sim}\num{e-2}$&109&\0\02&109&103   \\
Propyl formate & 1.89  &8.8 &0.0305&${\sim}\num{e-3}$&126&${\sim}
\num{e-1}$&126&126   \\
\bf{Acetates} \\
Methyl acetate & 1.72  &6.9 &0.0006&${\sim}\num{e-6}$&116&${\sim}
\num{e-4}$&116&122  \\
Ethyl acetate & 1.78  &8.6 &0.0048&${\sim}\num{e-4}$&156&${\sim}
\num{e-2}$&156&160  \\
\bf{Nitriles} \\
Acetonitrile & 3.93 &4.4& 1.175&27&155&202&182&180   \\
Propionitrile & 4.05  & 6.3&1.24&31&218&218&249&245   \\
2-methylpropionitrile  & 4.29  &8.1 &1.40&35&244&235&279&274   \\
\bf{Methyl halides} \\
Methyl fluoride & 1.86 & 2.4 & 0.042 & ${\sim}\num{e-4}$ & 0.33 & ${\sim}
\num{e-2}$ & 0.33 & 0.3 \\
Methyl chloride & 1.90 & 4.4 & 0.026 & ${\sim}\num{e-3}$ & 23 & ${\sim}\num{e-1}$ 
& 23 & 25 \\
Methyl bromide & 1.82 & 5.6 & 0.0085 & ${\sim}\num{e-3}$ & 42 & ${\sim}
\num{e-1}$ & 42 & 40 \\
\br
\end{tabular}
\end{table}

All of these new values of $r_0$ are significantly larger than the original
values. The three nitriles are the most strongly polar molecules studied, and
they now have very realistic values of $r_0$. The dipole in these nitriles is
due to the C$\equiv$N bond. The length of this bond is
\SI{2.19}{au} \cite{nist}, half of which is approximately \SI{1.1}{au}. The
values of $r_0$ are only slightly greater than this.

The ketones --- acetone, 2-butanone and cyclopentanone --- are the second most 
polar group. The polarity of these molecules is due to a C=O bond, the length
of which is \SI{2.26}{au} \cite{nist} (half of this is \SI{1.13}{au}). Their
values of $r_0$ are not as close to this estimate as those for the nitriles.
For the most polar molecule in the group, cyclopentanone, $r_0$ is within 19\%
of half the bond length.  The values of $r_0$ for acetone and 2-butanone are,
however, significantly smaller. The picture is similar for the aldehydes, which
show $r_0 \sim \SI{0.5}{au}$. The three other groups (formates, acetates and
methyl halides) have dipole moments $\mu\leq \SI{2}{D}$, only slightly exceeding
the critical dipole moment $\mu_{\rm crit} =\SI{1.625}{D}$. They all yield
unphysically small values of $r_0$.

The results suggest that our model, with the inclusion of polarizability via 
perturbation theory, is viable for molecules with dipole moments greater than
about \SI{3.5}{D}. It is of some concern that for all of the molecules studied,
including those for which we have now found realistic values of $r_0$, the
first-order energy corrections $\eb^{(1)}$ are much larger than the zeroth-order
energies $\eb^{(0)}$. However, one should compare the perturbative correction
with the mean potential energy in the original dipole potential
$\langle V_\mathrm{d} \rangle$, not the eigenvalue (in which the negative
potential energy and positive kinetic energy
contributions noticeably cancel each other). Table \ref{tab:perturbeddata}
shows this information for all of the molecules studied. We see that for the
most strongly polar molecules, e.g., the nitriles, $\langle V_\mathrm{d}\rangle$
and $\eb^{(1)}$ are of similar magnitude. This indicates that the corresponding
estimates of $\eb^{(1)}$ are reliable. On the other hand, for most of the other
molecules, the magnitude of $\langle V_\mathrm{d} \rangle$ is quite small
compared to that of $\eb^{(1)}$. However, even in these cases the
perturbation-theory estimate of the relative contribution of correlations
(i.e., polarization) appears to be sound, at least qualitatively.

It is interesting to compare the results from table \ref{tab:perturbeddata}
with real quantum chemistry calculations of positron binding to polar species.
The static dipole binding energy $\eb^{(0)}$ is then analogous to the static,
Hartree-Fock (HF) calculation of binding, while the total $\eb $ can be
compared with the configuration interaction (CI) result, which includes
correlations. In all cases the binding energy from the extensive CI
calculations is at least an order of magnitude greater than the HF value. For
example, for hydrogen cyanide (HCN, $\mu=\SI{3}{D}$), the binding energies are
\SI{1.6}{meV} (HF) and \SI{35}{meV} (CI) \cite{chojnacki}; for formaldehyde
(CH$_2$O, $\mu=\SI{3}{D}$), \SI{1.1}{meV} (HF) and \SI{19}{meV} (CI)
\cite{strasburger}; for nitrile molecules (CH$_3$CN, HCCCN, C$_2$H$_3$CN,
C$_2$H$_5$CN with $\mu=4.1$--\SI{4.4} {D}), the HF binding energies are
6--\SI{18}{meV}, becoming 81--\SI{164}{meV} in the CI calculation
\cite{tachikawa2}\footnote{In all likelihood the above CI energies
underestimate the true binding energy, because ``it is difficult to describe
the electron-positron correlation with the established methods of computational 
chemistry'' \cite{Wolcyrz2013}.}. The data for aldehydes, ketones and nitriles
in table \ref{tab:perturbeddata} show similar large increases due to the effect
of polarization. The model thus provides a useful estimate of the effect of
correlations on the binding energy.

As mentioned in the introduction, analysis of the measured binding energies 
found the dependence of $\eb$ on $\alpha$ for molecules within the same
chemical family (i.e., aldehydes, ketones, formates, acetates, nitriles) to be
almost linear \cite{danielson3}. From \eref{eqn:correction} we can see that,
for fixed $\mu$ and $r_0$, $\eb^{(1)}$ scales linearly with $\alpha$. Within each
chemical family, the type of dipole is the same and so $\mu$ does not vary much.
Thus, for the most part, the values of $r_0$ are fairly close to each other
within each chemical family.  This implies that considering the effect of
polarization as the first-order energy correction might be quite realistic,
even when $\alpha$ is large.

\subsection{Dependence of the binding energy on polarizability
for fixed $\mu$ and $r_0$}\label{subsec:BE_pol_pert}

\begin{figure}
\begin{indented}
\item[]\includegraphics*[width=.6\textwidth]{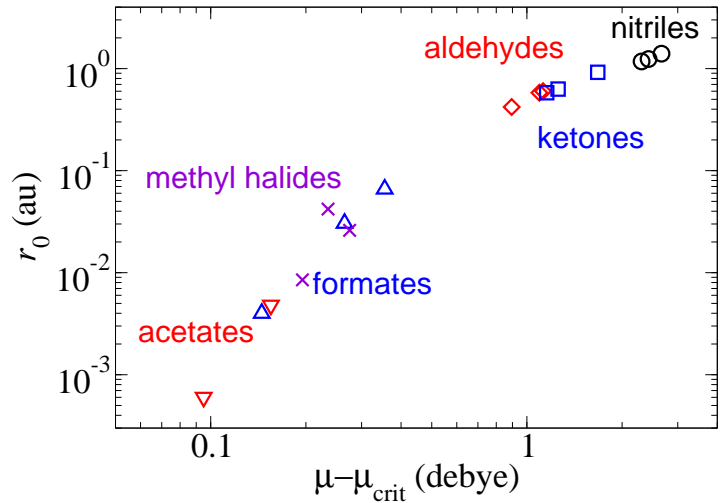}
\caption{Correlation between the molecular dipole moment and core radius $r_0$,
obtained from the perturbative polarization calculation
(see Table \ref{tab:perturbeddata}) for nitriles (circles), ketones (squares),
aldehydes (diamonds), formates (up triangles), acetates (down triangles), and
methyl halides (crosses).}
\label{fig:D_r0}
\end{indented}
\end{figure}

The new values of $r_0$ (those obtained after including the polarizability)
correlate strongly with the dipole moment of the molecules (see figure
\ref{fig:D_r0}). However, this correlation lacks an obvious physical basis, and
predicting the binding energy for an arbitrary molecule given only the values
of $\mu$ and $\alpha$ would be rather tenuous.

On the other hand, as was stated earlier, the dipole moment $\mu$ and core radius
$r_0$ do not change vastly from molecule to molecule within each chemical
family, for most of the families studied. To investigate the dependence
of $\eb$ on $\alpha$, we assigned to each family a fixed value of $\mu$ and
$r_0$. For five out of the six families, each with three molecules, we used the
values of $\mu$ and $r_0$ for the molecule with the median value of $\mu$. This
molecule will hereafter be referred to as the \emph{base molecule}. For the two
acetates, we arbitrarily chose ethyl acetate as the base molecule.

For the base molecule we know $\ed$ and $\ea$. By setting $\alpha$ to the 
appropriate values for the other molecules in the family, we were able to find
the corresponding $\ea$: they are just  linear rescalings of
\eref{eqn:correction}, since $\mu$ and $r_0$ had not changed. With $\ed$ fixed by
the values of $\mu$ and $r_0$ for the base molecule, we then had estimates of
$\eb$ for every molecule in the family.

\begin{figure}
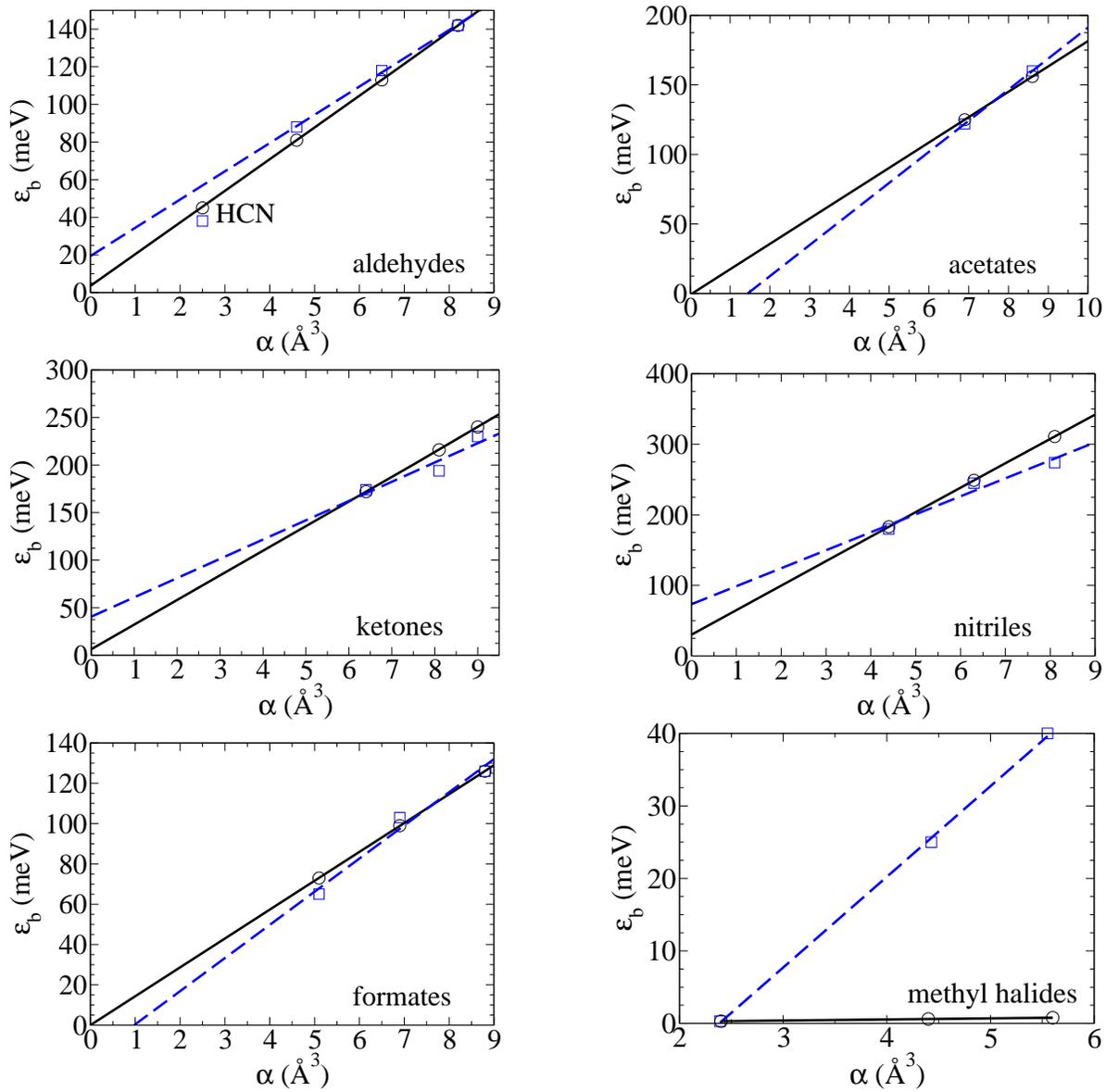

\includegraphics*[width=.45\textwidth]{figure5a.eps} \hfill
\includegraphics*[width=.45\textwidth]{figure5d.eps} \\
\includegraphics*[width=.45\textwidth]{figure5b.eps} \hfill
\includegraphics*[width=.45\textwidth]{figure5e.eps} \\
\includegraphics*[width=.45\textwidth]{figure5c.eps} \hfill
\includegraphics*[width=.45\textwidth]{figure5f.eps} 
\caption{\label{fig:extrapolation} Predicted and experimental/DMC values of
$\eb$ as functions of the dipole polarizability. The black circles and black,
solid line are the predicted values of $\eb$. The blue squares are the
experimental/DMC values of $\eb$, with the blue, dashed line a linear
regressive fit.}
\end{figure}

Table \ref{tab:extrapolation} compares the predicted and experimental values of
the binding energy for the seventeen molecules studied. We also use this method
to predict the binding energy of HCN, placing it in the aldehyde family
(see below). The results for each family are also shown graphically in figure
\ref{fig:extrapolation}. The dashed lines on the
graphs show linear fits of the measured binding energies, while the solids
lines display the linear dependence of the calculated binding energy on
$\alpha $, as described by equation \eref{eqn:correction}.

\begin{table}
\caption{\label{tab:extrapolation} The predicted values of $\eb$ found by using
fixed values of $\mu$ and $r_0$ for each chemical family, compared with the
experimental values. Hydrogen cyanide is included with the aldehydes as it has
a similar dipole moment; the value of $\eb$ for hydrogen cyanide obtained using
the diffusion Monte Carlo (DMC) method \cite{kita} is also given. The base
molecule for each family is indicated by `(base)' after its name.}
\lineup
\begin{indented}
\item[]\begin{tabular}{@{}llllll}
\br
&&&&\centre{2}{$\eb$ (meV)}  \\ \ns
&&&&\crule{2} \\
Molecule & $\mu$ (D) & $r_0$ (au) & $\alpha$ (\si{\angstrom^3}) & Pred. & Exp./DMC 
\\
\mr
\textbf{Aldehydes}  \\
Butanal (base) & 2.72 & 0.58 & 8.2 & 142 & 142 \\
Acetaldehyde & $''$ & $''$ & 4.6 & \081 & \088 \\
Propanal & $''$ & $''$ & 6.5 & 113 & 118 \\
{[Hydrogen cyanide]} & $''$ & $''$ & 2.5 & \045 & \038 \\
\textbf{Ketones} \\
Acetone (base) & 2.88 & 0.63 & 6.4 & 172 & 174 \\
2-butanone & $''$ & $''$ & 8.1 & 216 & 194 \\
Cyclopentanone & $''$ & $''$ & 9.0 & 240 & 230 \\
\textbf{Formates}  \\
Propyl formate (base) & 1.89 & 0.0305 & 8.8 & 126 & 126 \\
Methyl formate & $''$ & $''$ & 5.1 & \073 & \065 \\
Ethyl formate & $''$ & $''$ & 6.9 & \099 & 103 \\
\textbf{Acetates}  \\
Ethyl acetate (base) & 1.78 & 0.0048 & 8.6 & 156 & 160 \\
Methyl acetate & $''$ & $''$ & 6.9 & 125 & 122 \\
\textbf{Nitriles}  \\
Propionitrile (base) & 4.05 & 1.24 & 6.3 & 249 & 245 \\
Acetonitrile & $''$ & $''$ & 4.4 & 183 & 180 \\
2-methylpropionitrile & $''$ & $''$ & 8.1 & 311 & 274 \\ 
\textbf{Methyl halides}  \\
Methyl fluoride (base) & $1.86$ & $0.042$ & 2.4 & 0.33 & 0.3\\
Methyl chloride & $''$ & $''$ & 4.4 & 0.61 & 25 \\
Methyl bromide & $''$ & $''$ & 5.6 & 0.78 & 40 \\
\br
\end{tabular}
\end{indented}
\end{table}

For the aldehydes, the results are very good. In particular, note the similar
slopes of the experimental and predicted dependences of the binding energy on
$\alpha$. The predicted binding energies of acetaldehyde and propanal agree
with the experimental values to within 8\% and 5\% respectively.

For the ketones, the results are again pleasing. The predicted binding energies
of 2-butanone and cyclopentanone agree with the experimental values to within
12\% and 5\% respectively.

The formates also yield good results. The predicted binding energies of methyl 
formate and ethyl formate agree with the experimental values to within 13\% and
4\% respectively. It is actually quite surprising that the predicted binding
energies are as accurate as they are for this family, since $r_0$ for methyl
formate is an order of magnitude smaller than the values for the rest of the
family, and $r_0$ for ethyl formate is more than double the value for propyl
formate.

The predicted binding energy for methyl acetate is excellent; it is within 3\%
of the experimental value. Again, this is fairly surprising, given that $r_0$
for methyl acetate is an order of magnitude less than the value for ethyl
acetate. Note also that the $\alpha $ scaling holds well for both the formates
and acetates in spite of the unphysically small values of $r_0$.

Coming to the nitriles, we note the very encouraging results. The predicted
binding energies of acetonitrile and 2-methylpropionitrile agree with the
experimental values to within 2\% and 14\% respectively.

Finally, we note that for methyl halides the present model fails completely.
The predicted binding energies for $\mathrm{CH_3Cl}$  and $\mathrm{CH_3Br}$
are only 2\% of the measured values. At this point, we note that methyl halides
are the smallest molecules examined. The lightest of them, methyl fluoride, also
has the smallest moment of inertia, which means that molecular rotations
neglected by the model have the largest effect on this molecule. This, combined
with the smallest dipole polarizability, could be one of the reasons for the
anomalously small binding energy (0.3 meV) of this molecule. Hence, when the
``atypical'' $\mathrm{CH_3F}$ is chosen as the base molecule, the results for
the other two molecules are poor. Another reason that sets
methyl halides apart is that other molecules within each family consist of the
same types of atoms. They are quite similar chemically and have similar
ionization potentials (typically, not varying by more than \SI{0.5}{eV} within
each family \cite{CRC}). On the other hand, the three methyl halide molecules
contain different atoms (F, Cl or Br), and their ionization potentials vary
considerably more: \SI{12.47}{eV}, \SI{11.22}{eV} and \SI{10.54}{eV} for
$\mathrm{CH_3F}$, $\mathrm{CH_3Cl}$  and $\mathrm{CH_3Br}$ respectively 
\cite{CRC}.
This means that the additional attraction due to virtual positronium formation,
which is not accounted for by the dipole polarizability (see, e.g.,
\cite{DFGK95}), grows along this sequence. Since this additional attraction is
not present in our model, we obtain very poor predictions of binding energies.
At the same time, the dipole moment, even though not much greater than $\mu_{\rm crit}$,
plays a crucial role for binding by these molecules. Had these been nonpolar,
atom-like species, then, based on the their ionization potentials and dipole
polarizabilities, they would not have had bound states at all (see \cite{DFG00}
for the conditions of binding by atoms).

Overall, the perturbative treatment of polarization has been surprisingly
good for the five families of organic molecules. The maximum error in any of
the predicted binding energies is 14\%, even though two of the five families
exhibit small absolute values of $r_0$ with significant relative differences in
the values of $r_0$. The model also lends support to the empirical linear
relationship \eref{eqn:regression} between the binding energy and the dipole
polarizability,
even though the values of $r_0$ for most molecules are unphysically small.
In addition, it allows one to predict the binding energy for any 
molecule with a dipole moment that is comparable to those in one of the
chemical families studied, by placing it in that group and rescaling $\ea$ using
the appropriate polarizability.

As an example, consider hydrogen cyanide (HCN), which is a linear, triatomic 
molecule with a dipole moment of \SI{2.98}{D} and a polarizability of
\SI{2.5}{\angstrom^3} \cite{CRC}. Due to its toxicity, the
binding energy for hydrogen cyanide has not been measured experimentally.
Nevertheless, there already exist theoretical calculations of this energy using
a variety of methods, such as CI and diffusion Monte Carlo (DMC)
\cite{chojnacki,kita}. We estimated the binding energy by placing hydrogen
cyanide among the aldehydes, since they have similar dipole moments. The
resulting prediction of the binding energy (\SI{45}{meV}) is within 18\% of
the DMC value of \SI{38}{meV} \cite{kita} (see table \ref{tab:extrapolation}
and figure \ref{fig:extrapolation}).
Note that the ketones actually have more similar dipole moments to hydrogen 
cyanide than the aldehydes. Placing HCN in the ketone family led to a predicted 
binding energy of \SI{70}{meV}, which is a factor of two greater than the DMC
result of \cite{kita}. This large value is likely an overestimate, in spite of
the fact that quantum chemistry calculations tend to give lower bounds for the
positron binding energies \cite{tachikawa2,Tachikawa2012}. Experimental data
for ketones shows significant deviations from linearity (see figure
\ref{fig:extrapolation}), which makes the ketone-based prediction for HCN
less relaible.

\section{Non-perturbative treatment of molecular polarization}\label{sec:nonpert}

\subsection{Models of polarization potential and core radii}

Although the perturbative inclusion of polarization described in section
\ref{sec:perturb} generates reasonably accurate predictions of binding energies
for organic molecules, the effective core radii $r_0$ are too small for most
molecules to be physically meaningful. In addition, the first-order
polarization energy correction for most molecules is too large to justify the
use of perturbation theory. To overcome this limitation, in this section we
include the polarization potential in a full, non-perturbative manner. As we
will see, this leads to new physical insights and finally gives good physical
values of $r_0$ for all molecules.

When the polarization potential \eref{eq:Vpol} is added to the
Schr\"odinger equation \eref{eqn:schrodinger}, the angular equation
(\ref{eqn:angular}) remains unchanged, and we have the radial
equation
\begin{equation}\label{eqn:radialnonperturb}
-\frac{1}{2} \frac{\rmd^2P_{msn}}{\rmd r^2} + \left[ \frac{\lambda_{ms}}{2r^2}
-\frac{\alpha}{2r^4} g(r)\right] P_{msn}(r) = E_{msn} P_{msn}(r),
\end{equation}
for the function $P_{msn}(r)\equiv rR_{msn}(r)$. Here we have also
introduced the polarization cut-off function $g(r)$, which tends to unity at
large $r$ and moderates the unbounded, unphysical growth of the
$-\alpha /2r^4$ term at small distances (see, e.g., \cite{MBR02}; also see
below).

At large $r$, the polarization potential is negligible in comparison with the
$1/r^2$ dipole potential. Thus, at some sufficiently large value of
$r=r_\text{max}$ the radial wave function is given by equation \eref{eq:Rms},
which gives the boundary conditions
\numparts
\begin{eqnarray}
P_{msn}(r_\text{max}) &=&
\tilde A_{msn} \sqrt{r_\text{max}} K_{\rmi \beta_{ms}}(\kappa_{msn}r_\text{max}),\\
\left. \frac{\rmd P_{msn}(r)}{\rmd r} \right\vert_{r=r_\text{max}} &=&
\tilde A_{msn} \left.\frac{\rmd}{\rmd r}
\left[\sqrt{r}K_{\rmi \beta_{ms}}(\kappa_{msn}r)\right]\right\vert_{r=r_\text{max}},
\end{eqnarray}
\endnumparts
where $\tilde A_{msn}$ is an arbitrary constant. A value of
$r_\text{max}=\SI{30}{au}$ has been used throughout. Solving equation
(\ref{eqn:radialnonperturb}) numerically in the interval $0<r\leq r_\text{max}$
with $m=0$, $s=n=1$, $E_{011}=-\eb$, and $\tilde A_{011}=r_\text{max}^{-1/2}$
yields a real function $P_{011}(r)$ with infinitely many roots accumulating at $r=0$. The
largest of these roots is the value of $r_0$.

We initially considered $g(r)=1$, as in section \ref{sec:perturb}.
This led to values of $r_0$ in the range 1.55--2.32\,au across the six families
of molecules, which are much greater than their perturbative counterparts, and
probably too large to be considered physical. This is due to the polarization
potential \eref{eq:Vpol} blowing up and causing a rapid variation of
the radial wave function at small $r$, while in reality the polarization
is a long-range effect. For the same reason, the binding energy was found to be
extremely sensitive to the value of $r_0$, making it very difficult to use the
model in a predictive way.

Consequently, we considered several cut-off functions, viz.,
\numparts
\begin{eqnarray}
g_1(r) &=& 1 - \exp(-r^6/\rc^6), \\
g_2(r) &=& \frac{r^4}{(r+\rc)^4}, \\
g_3(r) &=& \frac{r^4}{(r^2+\rc^2)^2},
\end{eqnarray}
\endnumparts
where $\rc$ is a cut-off radius for the polarization. The function $g_1(r)$
provides a very rapid cut-off, and is commonly used to model polarization
potentials in atoms \cite{MBR02}. The functions $g_2(r)$ and $g_3(r)$
vary much more slowly. They can effectively account for the fact that the
``centre'' of the polarization potential is usually off-set with respect to
the location of the molecular dipole. The dipole moment is usually associated
with one of the terminal bonds, which is near one of the ``ends'' of the
molecule rather than in the middle. Initially we worked with fixed values of
$\rc$ across the entire
set of molecules, and though this reduced the values of $r_0$ from those of the
``hard'' potential ($g(r)=1$), and also reduced the sensitivity of the binding
energy to $r_0$, it did not significantly reduce the large spread in $r_0$
within or between families. This led us to consider using a
polarizability-dependent cut-off radius.

The polarizability of organic molecules is generally proportional to the number
of atoms or number of bonds in the molecule. This idea is the physical basis
behind various additivity methods for the calculation of molecular
polarizabilities \cite{miller}. In this spirit, the polarization
potential at large distances is the sum of terms $-\alpha _i/2r_i^4$ due to the
contribution of individual atoms or bonds $i$, with the distance $r_i$ measured
accordingly. In the spherically-averaged form \eref{eq:Vpol}, the distance
$r$ must measured from the ``centre of polarization'' rather than the centre
of the molecular dipole. As a result, at small $r$ the singular form
$-\alpha /2r^4$ must be replaced by a constant $-\alpha /2r_c^4$, as described
by the cut-off functions $g_2(r)$ and $g_3(r)$. Here $r_c$ is the effective
radius of the molecule. It is physical to link it to the polarizability
$\alpha$ by, e.g.,
\begin{equation}\label{eq:rc}
\rc = C \alpha^\mu,
\end{equation}
with $C$ being an adjustable parameter. The polarizability $\alpha$ has
dimensions of volume (i.e., length cubed), so the choice $\mu=\frac13$ would
be sensible for three-dimensional, approximately spherical molecules, while
$\mu=\frac12$ would be better for approximately planar molecules.
Experimentation showed that $\mu=\frac12$ works best for our set of molecules,
with $C$ being chosen separately for each family to minimize the range of
values of $r_0$ within the family.

\begin{figure}
\begin{indented}
\item[]\includegraphics*[width=.6\textwidth]{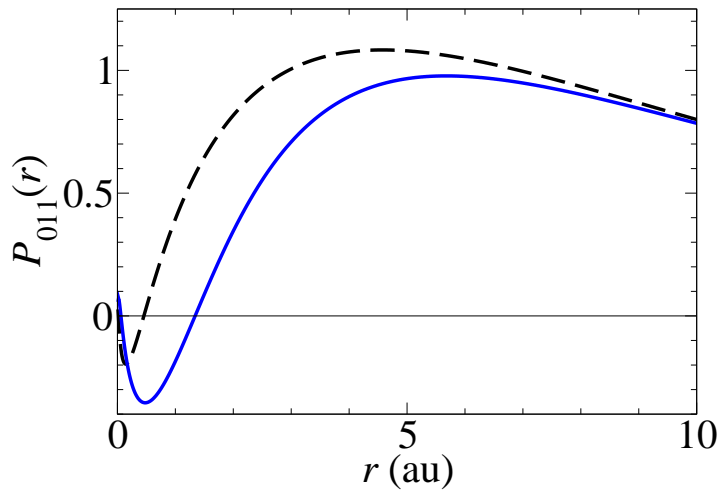} \hfill
\caption{\label{fig:acetonitrile_wfns} Radial wave functions for acetonitrile.
The dashed black curve is without the inclusion of polarization; the solid blue
curve is with the polarization included non-perturbatively, using the cut-off
function $g_3(r)$.}
\end{indented}
\end{figure}

Table \ref{tab:radii_nonperturb} shows the final values of $r_0$ obtained for
the molecules, with $\rc=C\alpha^{1/2}$. As expected, the cut-off functions
$g_2(r)$ and $g_3(r)$ gave the most physically meaningful results, with neither
being significantly better or worse than the other. Here we present the results
obtained using $g_3(r)$. Figure \ref{fig:acetonitrile_wfns} compares the radial
function $P_{011}(r)$ for acetonitrile, with and without the non-perturbative
inclusion of the polarization potential. It is clear from the figure that the
wave function does not change much for $r>\SI{5}{au}$. However, at smaller
distances the addition of the polarization potential causes a more rapid
variation of the wave function, leading to a greater core radius $r_0$.
\begin{table}
\caption{Values of $r_0$ when a soft polarization
potential with the cut-off function $g_3(r)$ is included non-perturbatively
and $r_c$ given by equation \eref{eq:rc} with $\mu =\frac12$. The parameter $C$
is given in units of $a_0$\,\si{\angstrom^{-3/2}}, where $a_0$ is the Bohr
radius.}
\label{tab:radii_nonperturb}
\lineup
\begin{indented}
\item[]\begin{tabular}{@{}lllll}
\br
Molecule & $\mu$ (D) &$\alpha$ (\si{\angstrom^3}) & $\eb$ (meV) &$r_0$ (au) \\
\mr
\textbf{Aldehydes} ($C=1.08$)\\
Acetaldehyde  & 2.75  &4.6& \088 & 1.17 \\
Propanal & 2.52  &6.5& 118 & 1.09 \\
Butanal & 2.72  &8.2& 142 & 1.16 \\
\textbf{Ketones} ($C=0.98$)\\
Acetone & 2.88  &6.4& 174 & 1.24 \\
2-butanone  & 2.78  &8.1& 194 & 1.24 \\
Cyclopentanone  & 3.30  &9.0& 230 & 1.37 \\
\textbf{Formates} ($C=1.06$)\\
Methyl formate  & 1.77  &5.1& \065 & 0.94 \\
Ethyl formate & 1.98  &6.9& 103 & 0.98 \\
Propyl formate & 1.89  &8.8& 126 & 0.94\\
\textbf{Acetates} ($C=0.96$)\\
Methyl acetate & 1.72  &6.9& 122 & 1.05 \\
Ethyl acetate & 1.78  &8.6& 160 & 1.05\\
\textbf{Nitriles} ($C=1.12$)\\
Acetonitrile & 3.93 &4.4& 180 & 1.34 \\
Propionitrile & 4.05  &6.3& 245 & 1.34\\
2-methylpropionitrile  & 4.29  &8.1& 274 & 1.43 \\
\textbf{Methyl halides} ($C=0.95$)\\
Methyl fluoride  & 1.86 &2.4& 0.3 & 1.22 \\
Methyl chloride  & 1.90 &4.4& 25 & 1.24 \\
Methyl bromide  & 1.82 &5.6& 40 & 1.24\\
\br
\end{tabular}
\end{indented}
\end{table}

As seen from table \ref{tab:radii_nonperturb}, all of the radii are now
${\sim}\SI{1}{au}$ and hence look physically meaningful. They also remain
approximately constant within each chemical family. The maximum range of $r_0$
within any family is \SI{0.13}{au} (for the ketones), and the range across
all the molecules is \SI{0.49}{au}. The values of $C$ are also quite consistent,
ranging from 0.95 to 1.12\,$a_0$\,\si{\angstrom^{-3/2}}. As before, the largest
radii are obtained for the most strongly polar molecules (the nitriles) and the
smallest radii are obtained for the most weakly polar molecules (the formates
and acetates).

\subsection{Dependence of the binding energy on polarizability
for fixed $\mu$ and $r_0$}

\begin{figure}[ht!]
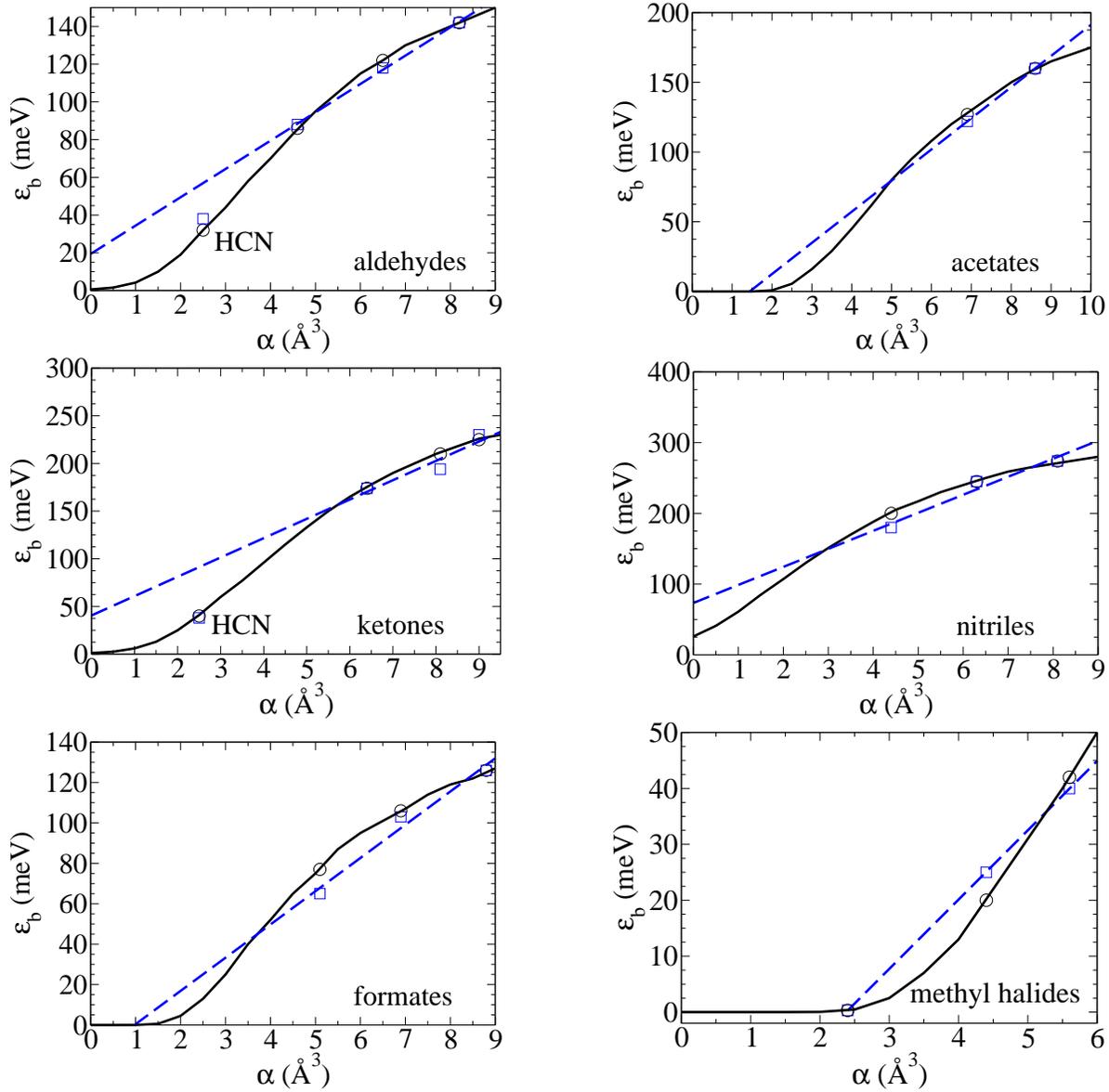

\includegraphics*[width=.45\textwidth]{figure7a.eps} \hfill
\includegraphics*[width=.45\textwidth]{figure7d.eps} \\
\includegraphics*[width=.45\textwidth]{figure7b.eps} \hfill
\includegraphics*[width=.45\textwidth]{figure7e.eps} \\
\includegraphics*[width=.45\textwidth]{figure7c.eps} \hfill
\includegraphics*[width=.45\textwidth]{figure7f.eps} 
\caption{\label{fig:binden_nonperturb} Predicted and experimental/DMC values of
$\eb$ as functions of the dipole polarizability, obtained using the
non-perturbative inclusion of the polarization potentil. The black circles are
the predicted values of $\eb$, with the black, solid curve showing the calculated dependence of $\eb$ on $\alpha$ for each family. The blue squares are the
experimental/DMC values of $ \eb$, with the blue, dashed line a linear
regressive fit.}
\end{figure}

We can now again investigate the dependence of the binding energy on the
molecular polarizability, by fixing $\mu$ and $r_0$ within each family and
varying $\alpha$. In these calculations we choose the same base molecule within
each family as in section \ref{subsec:BE_pol_pert}. The binding energy which
enters in equation \eref{eqn:radialnonperturb} is then adjusted for each
molecule in the family until the core radius $r_0$ of the base molecule is
obtained. Note that the polarization potential is now included
non-perturbatively, hence, there is no reason to expect that
$\eb $ depends lineary on $\alpha $. The resulting binding energies are shown
in table \ref{tab:binden_nonperturb} and figure \ref{fig:binden_nonperturb}.

\begin{table}[ht!]
\caption{\label{tab:binden_nonperturb} The predicted values of $\eb$ found by
using fixed values of $\mu$ and $r_0$ for each chemical family with the
polarization included non-perturbatively, compared with the experimental values.
Hydrogen cyanide is included with the aldehydes as it has a similar dipole 
moment; the value of $\eb$ for hydrogen cyanide obtained using the diffusion 
Monte Carlo method \cite{kita} is also given. The base molecule for each
family is indicated by `(base)' after its name.}
\lineup
\begin{indented}
\item[]\begin{tabular}{@{}llllll}
\br
&&&&\centre{2}{$\eb$ (meV)}  \\ \ns
&&&&\crule{2} \\
Molecule & $\mu$ (D) & $r_0$ (au) & $\alpha$ (\si{\angstrom^3}) & Pred. & Exp./DMC 
\\
\mr
\bf{Aldehydes} \\
Butanal (base) & 2.72 & 1.16 & 8.2 & 142 & 142 \\
Acetaldehyde & $''$ & $''$ & 4.6 & \086 & \088 \\
Propanal & $''$ & $''$ & 6.5 & 122 & 118 \\
{[Hydrogen cyanide]} & $''$ & $''$ & 2.5 & \032 & \038 \\
\bf{Ketones} \\
Acetone (base) & 2.88 & 1.24 & 6.4 & 174 & 174 \\
2-butanone & $''$ & $''$ & 8.1 & 210 & 194 \\
Cyclopentanone & $''$ & $''$ & 9.0 & 225 & 230 \\
\bf{Formates} \\
Propyl formate (base) & 1.89 & 0.94 & 8.8 & 126 & 126 \\
Methyl formate & $''$ & $''$ & 5.1 & \077 & \065 \\
Ethyl formate & $''$ & $''$ & 6.9 & 106 & 103 \\
\bf{Acetates} \\
Ethyl acetate (base) & 1.78 & 1.05 & 8.6 & 160 & 160 \\
Methyl acetate & $''$ & $''$ & 6.9 & 127 & 122 \\
\bf{Nitriles} \\
Propionitrile (base) & 4.05 & 1.34 & 6.3 & 245 & 245 \\
Acetonitrile & $''$ & $''$ & 4.4 & 200 & 180 \\
2-methylpropionitrile & $''$ & $''$ & 8.1 & 274 & 274 \\ 
\bf{Methyl halides}\\
Methyl fluoride (base) & $1.86$ & $1.24$ & 2.4 & 0.3 & 0.3\\
Methyl chloride & $''$ & $''$ & 4.4 & 20 & 25 \\
Methyl bromide & $''$ & $''$ & 5.6 & 42 & 40 \\
\br
\end{tabular}
\end{indented}
\end{table}

There is generally very close agreement between the model predictions and the
experimental data; for every molecule except methyl formate and methyl acetate,
the relative difference of the predicted binding energy from the experimental
value is smaller than it was using the perturbative method. Particularly
noteworthy is 2-methylpropionitrile, for which the predicted binding energy
coincides exactly with the experimental value. The error for methyl formate has
increased from 13\% to 18\%, and for methyl acetate it has increased from 3\%
to 4\%. 

From figure \ref{fig:binden_nonperturb} it is apparent that when the
polarization
potential is included non-perturbatively, the dependence of $\eb$ on $\alpha$ is
indeed non-linear. For all families except the methyl halides, $\eb$ increases
convexly with $\alpha$ for $\alpha \lesssim\SI{4}{\angstrom^3}$; for 
$\alpha \gtrsim \SI{4}{\angstrom^3}$ the growth becomes concave. The
molecules studied all lie in the convcave region, and the growth for them
could be reasonably well approximated by a straight line. The prediction curve
for
the methyl halides is different but particularly remarkable, as the description 
of this molecular family was extremely poor in the perturbative treatment.
Here the molecules lie in the convex region (which spans a larger range of
polarizabilities than for the other families), and the dependence of $\eb$ on
$\alpha$ is markedly nonlinear. However, close agreement is observed with the
measured binding energies for $\mathrm{CH_3Cl}$ and $\mathrm{CH_3Br}$.

Table \ref{tab:binden_nonperturb} and figure \ref{fig:binden_nonperturb} also
show our estimate of the binding energy for HCN. As seen from the graph, 
the experimental value lies very close to the prediction curve for the aldehydes data,
which provides support that the dependence of $\eb $ on $\alpha $ is not truly
linear. The value obtained (\SI{32}{meV}) is within 16\% of the DMC calculation
(\SI{38}{meV}) \cite{kita}, which is slightly closer than our perturbative
estimate. Interestingly, if we now place HCN in the ketones family, its
estimated binding energy becomes \SI{40}{meV}, in very close accord with the DMC
value. This is further evidence that a nonperturbative treatment of
molecular polarization gives overall much more consistent results.

\section{Concluding remarks}

Here we provided a simple model for positron binding to polar molecules, which 
captures the essential physics of this system.

Modelling the molecule as a sphere of radius $r_0$ with a static point dipole
of dipole moment $\mu$ at the centre and using experimental data on binding
energies required unphysically small values of $r_0$, even for the most
strongly polar molecules. This indicated that the binding energies are greatly
enhanced by some factor other than the molecule's permanent dipole moment,
i.e., electron-positron correlations.

Including the effect of correlations perturbatively through the polarization
potential did confirm this expectation. It showed that even for the strongly
polar molecules, the effect of correlations increased the binding energy by an
order of magnitude compared to the static-dipole calculation. The observed
increase matched the difference between the CI and HF binding energies obtained
in state-of-the-art quantum chemistry calculations.

Including the polarization potential as a perturbation of the original
Hamiltonian also yielded larger, more physical values of $r_0$ for all of the
molecules studied, but for most molecules they were still too small to be
interpreted directly. This was partly due to the fact that the true static
potential for the positron near the molecule is less repulsive than the hard
wall of our model. Reduced values of $r_0$ may also account for some of the
short-range correlation effects, such as virtual positronium formation.
Sensible values of $r_0$ were, however, obtained for the nitriles, the most
strongly polar of all the molecules studied. In spite of the fact
that most of the molecules had unphysical values of $r_0$, it was found that
taking the value of $\mu$ and $r_0$ for the molecule in each chemical family with
the median dipole moment and varying the polarizability to match the other
molecules in the family, gave reliable predictions of the binding energies
for those molecules (with the exception of the methyl halides). The perturbative
treatment was also in line with the observation made by experimentalists,
that the dependence of the binding energy on the polarizability of the molecule
is apparently almost linear \cite{danielson3}. Of course, a general increase in
the binding energy with the polarizability could be expected, but there was no
explanation for the linear dependence. According to our model, this feature
indicates that the perturbation theory is at least qualitatively correct, even
though the first-order energy corrections are generally greater than the
original (dipole) eigenenergies. The results of this treatment for the methyl
halides, however, were very poor in comparison with the other families, and we
attributed this to the fact that the binding by the base molecule
($\mathrm{CH_3F}$) is likely affected by rotations, and that the three methyl
halide molecules had a significantly larger range of ionization potentials
that the other families. The latter could indicate a significant change in the
contribution of virtual positronium formation across the members of this
family, which is not accounted for in our model.

A test of the model was to use it to predict the binding energy for hydrogen 
cyanide, which has not been measured experimentally. Our estimate agreed with a 
previous calculation using the diffusion Monte Carlo method to within 20\%,
which provides evidence that our model has good predictive power and could
be useful for estimating the binding energies that have never been measured in 
experiment, provided that the binding energy for a molecule with a similar
value of $\mu$ is known.

The most glaring limitation of the model as it stood was that it could not
predict binding energies using only the dipole moment and polarizability of a
molecule. To perform a calculation one needed a value of $r_0$, and unless
binding energies for molecules with similar values of $\mu$ were known, one could
not easily choose a suitable value for $r_0$. In fact, we found that for most
of the chemical families we considered the values of $r_0$ had no immediate
physical relevance, and for the most weakly polar families (i.e., formates and
acetates) there were significant variations in the values of $r_0$ despite the
similar values of $\mu$.

In a bid to attain physically meaningful values of $r_0$ for all of the
molecules, we proceeded to include the effects of polarization in a
non-perturbative way. We experimented by solving the radial Schr\"odinger
equation numerically using several model polarization potentials, and found
that the best results were obtained using a polarization potential
of the form $-\alpha/2(r^2+\rc^2)^2$, with the cut-off radius
$\rc=C\alpha^{1/2}$ ($C$ being a constant). The parameter $C$ was chosen
separately for each chemical family so as to minimize the spread of $r_0$
within each family. Physically meaningful values of $r_0$ (in the range
0.94--1.43\,au) were then obtained for all molecules. By choosing the values of
$C$ carefully, the spread of values of $r_0$ within in each family was made
relatively small, though there was inevitably still a larger range of
\SI{0.49}{au} across the entire set of molecules. The most strongly (weakly)
polar molecules still possessed the largest (smallest) values of $r_0$. Again
fixing $\mu$ and $r_0$ for each family and varying only the polarizability led to
excellent predictions of $\eb$; the predictions were generally more accurate
than their perturbative counterparts, with particularly large improvement for
the methyl halides. We observed that the true dependence of $\eb$ on
$\alpha$ is actually nonlinear.
The prediction for HCN was also slightly better than its
perturbative counterpart, and supported our observation of nonlinear growth
of $\eb$ with $\alpha$.




In summary, our model can be used effectively to predict positron-molecule
binding energies based on the molecular dipole moment and dipole polarizability,
particularly when polarization is included in a non-perturbative way. 
It provides a clear picture of the system, thereby complementing the current
computational effort towards rigorous theory of positron-molecule binding.

\ack
The authors are grateful to I I Fabrikant for useful discussions and
suggestions.

\section*{References}


\providecommand{\newblock}{}

\end{document}